\documentclass[aps,prd,twocolumn,showpacs,nofootinbib]{revtex4-2}

\usepackage{graphicx}
\usepackage{float}
\usepackage{color}
\usepackage{slashed}
\usepackage[colorlinks=true,linkcolor=blue,citecolor=blue, urlcolor=blue]{hyperref}

\begin{document}

\title{EEXICC: An event generator for doubly heavy baryon production at $e^+e^-$ colliders}

\author{Hong-Tai Li$^{a}$}
\email{liht@cqu.edu.cn}
\author{Xu-Chang Zheng$^{a}$}
\email{zhengxc@cqu.edu.cn}
\author{Xing-Gang Wu$^{a}$}
\email{wuxg@cqu.edu.cn}
\author{Jiang Yan$^{b}$}
\email{yjiang@itp.ac.cn}
\author{Zhi Yang$^{c,d}$}
\email{zhiyang@uestc.edu.cn}

\address{$^{a}$ Department of Physics, Chongqing Key Laboratory for Strongly Coupled Physics, Chongqing University, Chongqing 401331, People's Republic of China\\
$^b$ Institute of Theoretical Physics, Chinese Academy of Sciences, Beijing 100049, P.R. China\\
$^c$ Yangtze Delta Region Institute (Huzhou), University of Electronic Science and Technology of China, Huzhou 313001, People's Republic of China\\
$^d$ School of Physics, University of Electronic Science and Technology of China, Chengdu 610054, People's Republic of China}

\begin{abstract}

We present EEXICC, a Monte Carlo event generator designed to simulate the production of doubly heavy baryons ($\Xi_{cc}$, $\Xi_{bc}$, and $\Xi_{bb}$) via $e^+e^-$ annihilation. Based on nonrelativistic QCD effective theory, the generator calculates the process $e^{+}+e^{-}\rightarrow \Xi_{QQ'}+\bar{Q}'+\bar{Q}$ using an improved trace technique at the amplitude level, which greatly improves numerical efficiency compared with traditional squared-amplitude methods. EEXICC uses a modular Fortran implementation for hard-process generation and a C++ interface to the bundled PYTHIA 8.317 library, enabling convenient integration into complete event simulation workflows. The program supports both weighted and unweighted event generation, and its numerical reliability has been verified against existing theoretical results. EEXICC provides a flexible and robust tool for studying the properties of doubly heavy baryons at future high-luminosity and high-energy $e^+e^-$ colliders such as the CEPC and FCC-ee. \\



\end{abstract}

\maketitle

\noindent{\bf Program summary}\\

\noindent{\it Title of program}: EEXICC\\

\noindent{\it Version}: 1.0 \\

\noindent{\it Program obtained from}: CPC Program Library.\\

\noindent{\it Computer}: Any workstation or cluster with a Linux operating system.\\

\noindent{\it Operating systems}: Linux.\\

\noindent{\it Programming language used}: Fortran 77/90, C++.\\

\noindent{\it Memory required to execute with typical data}: $\sim$ 256 MB.\\

\noindent{\it Distribution format}: Compressed tar file.\\

\noindent{\it External routines/libraries}: PYTHIA 8.317 (required by the current build); GNU Fortran, a C++11 compiler and GNU Make.\\

\noindent{\it Nature of problem}: The program simulates the production of doubly heavy baryons ($\Xi_{cc}$, $\Xi_{bc}$, and $\Xi_{bb}$) \footnote{In this paper, we use $\Xi_{QQ'}$ to denote a generic doubly heavy baryon containing two heavy quarks, $Q$ and $Q'$.} via $e^+e^-$ annihilation at high-energy colliders. It is designed to investigate the properties and production rates of these baryons for future experiments.\\

\noindent{\it Solution method}: The simulation is based on the nonrelativistic QCD factorization framework. The program calculates the production amplitudes using an improved trace technique, which enhances numerical stability and speed. Both color $\mathbf{\bar{3}}$ and color $\mathbf{6}$ $S$-wave $(QQ')$ intermediate states are included. The generator offers options for both weighted and unweighted events, where unweighted event generation is optimized using an improved hit-and-miss algorithm to maximize efficiency.\\

\noindent{\it Additional comments including restrictions and unusual features}: The current version implements a leading-order calculation in both the strong coupling constant and the typical heavy-quark velocity in the baryon rest frame, based on the nonrelativistic QCD framework; consequently, neither QCD radiative corrections nor relativistic corrections are included. Furthermore, the generator focuses exclusively on the $e^+e^-$ annihilation channel. Contributions from photon-photon scattering processes, which become significant at collision energies far above the $Z$-boson resonance, are not currently implemented. \\

\section{background and main idea of the generator EEXICC}

Doubly heavy baryons, consisting of two heavy quarks and one light quark, are particularly fascinating as they provide a fresh perspective on the problems of heavy-quark production and hadronization. Compared to ordinary heavy baryons, doubly heavy baryons involve more distinct scales, such as the heavy-quark mass ($m_Q$), the typical heavy-quark momentum ($m_Q v$), and the nonperturbative QCD scale ($\Lambda_{\rm QCD}$). Consequently, theoretical predictions for doubly heavy baryons are more challenging, but they can reveal the properties of QCD from diverse perspectives. In 2017, the LHCb collaboration reported the observation of the doubly charmed baryon $\Xi_{cc}^{++}$ via the decay $\Xi_{cc}^{++} \to \Lambda_c^+ K^- \pi^+ \pi^+$ (with $\Lambda_c^+ \to p K^- \pi^+$)~\cite{LHCb:2017iph}, and this discovery was subsequently confirmed through the observation of the decay $\Xi_{cc}^{++} \to \Xi_{c}^{+} \pi^+$~\cite{LHCb:2018pcs}, followed by the observation of other decay modes~\cite{LHCb:2022rpd}. Very recently, the LHCb collaboration announced the first observation of $\Xi_{cc}^{+}$ via the decay $\Xi_{cc}^{+} \to \Lambda_c^+ K^- \pi^+$ using the LHCb Run 3 detector~\cite{LHCb:2026pxn, LHCb:2025shu}. The experimental observation of doubly charmed baryons has revitalized interest in both theoretical and experimental studies of doubly heavy baryons. In recent years, numerous theoretical studies on the direct and indirect production of doubly heavy baryons at various high-energy colliders have been performed~\cite{Bi:2017nzv, Sun:2020mvl, Chen:2019ykv, Chen:2018koh, Koshkarev:2016acq, Koshkarev:2016rci, Groote:2017szb, Berezhnoy:2018bde, Berezhnoy:2018krl, Niu:2018ycb, Niu:2019xuq, Zhang:2022jst, Luo:2022jxq, Luo:2022lcj, Li:2020ggh, Ma:2022cgt, Ma:2022ger, Niu:2023ojf, Ma:2025ito}.

An event generator for simulating doubly heavy baryon production at high-energy colliders is highly beneficial for both theoretical and experimental studies of doubly heavy baryons. In 2007, a dedicated and highly effective generator named GENXICC~\cite{Chang:2007pp, Chang:2009va}, implemented in a PYTHIA-compatible format~\cite{Sjostrand:2006za}, was developed to simulate doubly heavy baryon production at hadron colliders. GENXICC has since become a widely adopted tool for such simulations. Nevertheless, the complex environment at hadron colliders limits the precision of experimental studies on doubly heavy baryons. Therefore, it is worthwhile to investigate the properties of doubly heavy baryons at alternative experimental platforms. 

Electron–positron ($e^+e^-$) colliders provide an important complement to hadron colliders, benefiting from a much cleaner experimental environment and well-defined initial states. In the literature, several next-generation $e^+e^-$ collider projects are currently under active consideration, including the Circular Electron–Positron Collider (CEPC)~\cite{CEPCStudyGroup:2023quu}, the Future $e^+e^-$ Circular Collider (FCC-ee)~\cite{FCC:2018evy}, the International Linear Collider (ILC)~\cite{ILCInternationalDevelopmentTeam:2022izu}, and the Super $Z$ Factory~\cite{Z-factory}. These facilities are expected to operate at the $Z$-boson resonance for extended periods, achieving luminosities that exceed those of LEP-I by three to four orders of magnitude. It has been shown that the production rates of doubly heavy baryons are greatly enhanced at center-of-mass energies around the $Z$ pole~\cite{Jiang:2012jt, Jiang:2013ej}. Consequently, future high-luminosity $e^+e^-$ colliders offer a unique opportunity to investigate the production mechanisms and properties of doubly heavy baryons with unprecedented precision. Motivated by this prospect, we develop an event generator, referred to as EEXICC, for the simulation of doubly heavy baryon production in $e^+e^-$ collisions.

\begin{figure}[htb]
\includegraphics[width=0.48\textwidth]{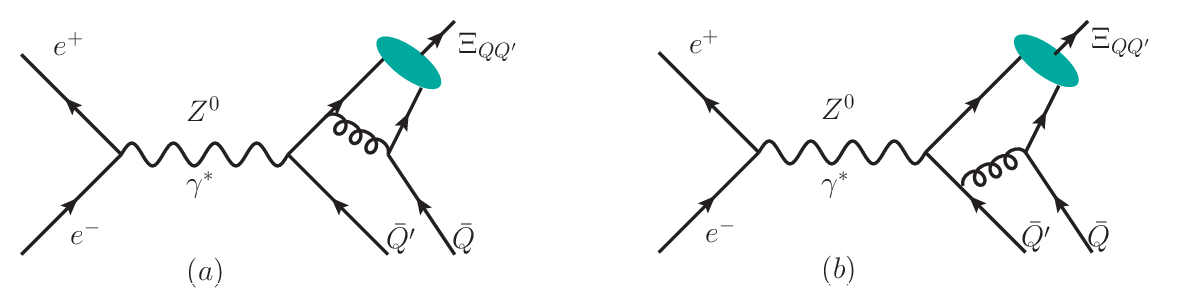}
\caption{Typical Feynman diagrams for the production of doubly heavy baryons via $e^+e^-$ annihilation, i.e., $e^+ + e^- \to \Xi_{QQ'} + \bar{Q} + \bar{Q}'$, where $Q$ and $Q'$ denote $b$ or $c$ quarks.} \label{feyfig}
\end{figure}

In the framework of the nonrelativistic QCD (NRQCD) effective theory~\cite{nrqcd}, the production process of a doubly heavy baryon (e.g., $\Xi_{QQ'}$) can be factorized into two stages: the first stage is the production of a heavy quark pair ($QQ'$) with a ``small relative velocity" and proper quantum numbers, and the second stage is the hadronization of the $QQ'$ pair into the doubly heavy baryon $\Xi_{QQ'}$. Here, $Q$ and $Q'$ denote $b$ or $c$ quarks. The production rate for the first stage can be calculated using perturbative QCD, since the relevant scales in this stage are comparable to or larger than the heavy quark mass. The second stage involves the $QQ'$ pair combining with the light QCD degrees of freedom to form the doubly heavy baryon $\Xi_{QQ'}$. The transition probability of the second stage can be described by the nonperturbative NRQCD matrix element. For the $Q \neq Q'$ case, there are four $Z$-boson-exchange Feynman diagrams and four photon-exchange Feynman diagrams. Two of the $Z$-boson/photon-exchange diagrams are presented in Fig.~\ref{feyfig}, and the remaining two can be obtained by interchanging the fermion lines of $Q$ and $Q'$ in Fig.~\ref{feyfig}. For the $Q = Q'$ case, there are eight $Z$-boson-exchange Feynman diagrams and eight photon-exchange Feynman diagrams. Besides the two $Z$-boson/photon-exchange diagrams shown in Fig.~\ref{feyfig}, the other six diagrams can be obtained by interchanging the positions of the two identical quarks or the two identical antiquarks in Fig.~\ref{feyfig}.

Since the production processes of doubly heavy baryons involve the generation of a heavy quark pair $(QQ')$, the calculation for these processes is rather complicated and lengthy. To improve the efficiency of the numerical simulation, we adopt the ``improved trace technique" proposed in Ref.~\cite{Chang:1992bb}, in which the relevant processes are treated directly at the amplitude level. After generating appropriate phase-space points, the numerical values of the amplitudes are first calculated, and these values are then algebraically summed and squared to obtain the squared amplitude, i.e., $|{\cal M}|^2=|\sum_{i}{\cal M}_{i}|^2$. In this manner, the efficiency of the numerical simulation can be significantly improved compared with the conventional squared-amplitude technique. Furthermore, owing to the symmetry of the fermion lines and the properties of various heavy-quark states, considerable simplifications can be made at the amplitude level. Indeed, the analytical expressions for the amplitudes can even be written as a sum of several independent Lorentz structures, each with corresponding coefficients. A detailed description of the improved trace technique can be found in Refs.~\cite{Chang:2007si, Deng:2010aq, Yang:2010yg, Yang:2011ps}. All the independent Lorentz structures and their coefficients are implemented in the EEXICC generator as separate subroutines.

The EEXICC package is developed in the PYTHIA format to ensure that it can be easily adopted into the PYTHIA environment for complete event generation. For the phase-space integration, the subroutine RAMBOS~\cite{Kleiss:1985gy} is adopted to generate the necessary phase-space points. To improve the Monte Carlo simulation efficiency for high-dimensional phase-space integration, EEXICC includes a switch to select whether to use the VEGAS program~\cite{Lepage:1977sw} to obtain the sampling importance function. The package also implements key parameters, including the maximum differential cross section, to satisfy the initialization requirements of PYTHIA.

To verify the correctness of the EEXICC program, we reproduce the calculations in Ref.~\cite{Zheng:2015ixa}. Using the same input parameters, we numerically compute the total cross section for the process $e^{+}e^{-}\to \Xi_{QQ'}+\bar{Q}+\bar{Q'}$ by integrating over the unobserved variables, and then carefully compare our results with those presented in Ref.~\cite{Zheng:2015ixa}. This check is highly nontrivial because EEXICC employs an improved trace technique that differs from the approach adopted in Ref.~\cite{Zheng:2015ixa}.

The paper is organized as follows. In Sec. II, we outline the main procedures of the calculation methods for handling $\Xi_{QQ'}$ production. In Sec. III, we present the key features of the EEXICC generator, including detailed descriptions of its structure, flow chart, and usage. The final section is dedicated to a summary.

\section{Calculation methods}

According to the NRQCD factorization~\cite{nrqcd}, the differential cross section for the process $e^{-}(q_1)+e^{+}(q_2) \rightarrow \Xi_{QQ'}(p_1) +\bar{Q}(p_2) +\bar{Q'}(p_3)$ can be expressed as:
\begin{eqnarray}
&& d\sigma(e^{+}e^{-}\rightarrow \Xi_{QQ'} +\bar{Q} +\bar{Q'}) \nonumber \\
&& = \sum_{n} d\hat\sigma (e^{+}e^{-}\rightarrow (QQ')[n] +\bar{Q} +\bar{Q'}) \langle{\cal O}^H(n)\rangle,
\label{eq.nrqcd}
\end{eqnarray}
where $d\hat\sigma$ represents the short-distance coefficient (SDC), $\langle{\cal O}^H(n)\rangle$ is the nonperturbative long-distance matrix element (LDME), and $n$ denotes the quantum numbers of the intermediate $(QQ')$ pair. As discussed in Ref.~\cite{Ma:2003zk}, at leading order in $v_Q$ (or $v_{Q'}$), which represents the typical heavy-quark velocity in the baryon rest frame, the $(QQ')$ pair can be in either a color $\overline{\mathbf{3}}$ or a color $\mathbf{6}$ $S$-wave state. Constraints from Fermi-Dirac statistics require the wave function of the pair to be totally antisymmetric under the exchange of identical quarks. Consequently, for the identical heavy-quark case ($Q=Q'$), the allowed states for $n$ are restricted to $^3S_1^{[\overline{\mathbf{3}}]}$ and $^1S_0^{[\mathbf{6}]}$; whereas for the distinct heavy-quark case ($Q\neq Q'$), $n$ can be $^1S_0^{[\overline{\mathbf{3}}]}$, $^3S_1^{[\overline{\mathbf{3}}]}$, $^1S_0^{[\mathbf{6}]}$, or $^3S_1^{[\mathbf{6}]}$, respectively.

The SDC $d\hat\sigma(e^{+}+e^{-}\rightarrow (QQ')[n]+\bar{Q}+\bar{Q'})$ represents the differential cross section for the production of the $(QQ')$ pair, which can be calculated through perturbative QCD. More explicitly, the SDC can be written as
\begin{eqnarray}
d\hat\sigma(e^{+}e^{-}\to (QQ')[n]+\bar{Q}+\bar{Q'})= \frac{\overline{\sum}  |{\cal M}|^{2} d\Phi_3}{4\sqrt{(q_1\cdot q_2)^2-m_e^4}} ,
\end{eqnarray}
where $m_e$ is the electron mass, $1/(4\sqrt{(q_1\cdot q_2)^2-m_e^4})$ is the flux factor, $\overline{\sum}$ means the average over the spin states of initial particles and the sum over the color and spin states of all final particles. The three-particle differential phase space is
\begin{displaymath}
d{\Phi_3}=(2\pi)^4 \delta^{4}\left(q_1+q_2 - \sum_f^3 p_{f}\right)\prod_{f=1}^3
\frac{d^3{p_f}}{(2\pi)^3 2p_f^0}.
\end{displaymath}
If $Q'=Q$, the cross section should be multiplied by an additional factor $(1/2!)^2$ due to the presence of identical quarks and identical antiquarks in the final states.

The scattering amplitudes ($\mathcal{M}$) for $(QQ')$ pair production can be directly obtained from the Feynman diagrams. Before evaluating the amplitudes, we first reverse one fermion line in each Feynman diagram (e.g., the $Q'$ fermion line) by inserting proper charge-conjugation operators. The amplitudes for $(QQ')$ production can then be expressed in terms of those for $(Q\bar{Q}')$ production. For the $(Q\bar{Q}')$ production amplitudes, the spin projectors used for meson production can be applied. Further details of the fermion-line reversal approach can be found in Refs.~\cite{Chang:2006eu, Zheng:2015ixa}. To further simplify the amplitude calculations, we adopt the ``improved trace technique" proposed in Ref.~\cite{Chang:1992bb}. This technique directly evaluates the trace at the amplitude level and expresses the amplitudes in terms of Lorentz structures. This method greatly reduces computational time compared with the conventional squared-amplitude technique.

The LDMEs describe the transition probabilities for heavy quark pairs into doubly heavy baryons, which are universal and can be extracted from the experimental data. However, the current experimental data on doubly heavy baryons are insufficient to extract them. The LDMEs for the color $\overline{\bf 3}$ states can be estimated through the wave functions of the color $\overline{\bf 3}$ diquark. More explicitly, the transition of a $(QQ')$ pair in the color $\overline{\bf 3}$ state to the doubly heavy baryon $\Xi_{QQ'}$ can be divided into two steps: the formation of bound diquark in the color $\overline{\bf 3}$ state that can be described by the wave function of the color $\overline{\bf 3}$ diquark, and the fragmentation of the diquark into $\Xi_{QQ'}$. Since the color $\overline{\bf 3}$ diquark can combine a light quark from the collision environment easily, the probability of the fragmentation from the color $\overline{\bf 3}$ diquark into $\Xi_{QQ'}$ is usually assumed as $\approx 100\%$. Therefore, the transition probability of a $(QQ')$ pair in the color $\overline{\bf 3}$ state to $\Xi_{QQ'}$ can be estimated by the wave function at origin of the color $\overline{\bf 3}$ diquark, i.e., 
\begin{eqnarray}
\langle{\cal O}^{\Xi_{QQ'}}(^{3(1)}S_{1(0)}^{[\overline{\bf 3}]})\rangle \approx \vert \Psi_{QQ'}(0)\vert^2=\frac{\vert R_{QQ'}(0)\vert^2}{4\pi}.
\label{eq.LDME3}
\end{eqnarray}
However, there is no analogous relation for the color-${\bf 6}$ LDMEs. Following the discussion in Ref.~\cite{Ma:2003zk}, the color ${\bf 6}$ LDMEs are of the same order in $v_{Q}(v_{Q'})$ as the corresponding $\overline{\bf 3}$ LDMEs. Consequently, we take the color ${\bf 6}$ LDMEs as the same values as the corresponding color $\overline{\bf 3}$ LDMEs to make our estimates, i.e., 
\begin{eqnarray}
\langle{\cal O}^{\Xi_{QQ'}}(^{3(1)}S_{1(0)}^{[{\bf 6}]})\rangle \approx \langle{\cal O}^{\Xi_{QQ'}}(^{3(1)}S_{1(0)}^{[\overline{\bf 3}]})\rangle.
\label{eq.LDME6}
\end{eqnarray}
Fortunately, if more precise determinations of these LDMEs become available, their values can be conveniently updated in the generator.

Eq.~(\ref{eq.nrqcd}) gives the total cross sections for the production of doubly heavy baryons containing a $(QQ')$ pair, summed over all possible light constituent quark flavors ($u$, $d$, and $s$). However, doubly heavy baryons with different light quarks are experimentally distinguishable. Therefore, providing predictions for these specific states is of great phenomenological importance. The relative production probabilities can be estimated within the Lund string model~\cite{Sjostrand:2006za}, which suggests a ratio $\sigma(\Xi_{QQ'u}):\sigma(\Xi_{QQ'd}):\sigma(\Xi_{QQ's}) \approx 1:1:0.3$. In this generator, we calculate the cross sections for specific baryons by multiplying the results from Eq.~(\ref{eq.nrqcd}) by the corresponding fragmentation fractions. For example, for doubly charmed baryons, we obtain $\sigma(\Xi_{cc}^{++}) \approx \sigma(\Xi_{cc}^{+}) \approx 43.5\% \, \sigma(\Xi_{cc})$ and $\sigma(\Omega_{cc}^{+}) \approx 13.0\% \, \sigma(\Xi_{cc})$.

To compute the total and differential cross sections for the relevant processes, we first use the RAMBOS subroutine~\cite{Kleiss:1985gy} to generate the phase-space points. This subroutine transforms the four-momenta of the outgoing particles into a convenient form for evaluating the scattering amplitudes $\mathcal{M}$ and their squared moduli $|\mathcal{M}|^2$. For the numerical integration over the multi-dimensional phase space, which is essential for obtaining physical cross sections, we employ the adaptive Monte Carlo integration program VEGAS~\cite{Lepage:1977sw}. VEGAS is particularly efficient for this task because it iteratively optimizes the sampling distribution to enhance the precision of the integral, making it highly suitable for high-dimensional quantum field theory calculations.

\section{The generator EEXICC}

The EEXICC generator is specifically designed to simulate the production of doubly heavy baryons ($\Xi_{cc}$, $\Xi_{bc}$, and $\Xi_{bb}$) at $e^+e^-$ colliders. The hard process is generated in Fortran and can be passed in memory to a C++ adapter linked to PYTHIA 8.317 \cite{Bierlich:2022pfr}. This adapter supports a restricted hadronization workflow, rather than unrestricted access to all PYTHIA simulation stages.

\subsection{Structure of EEXICC}

\begin{figure}[htb]
\includegraphics[width=0.48\textwidth]{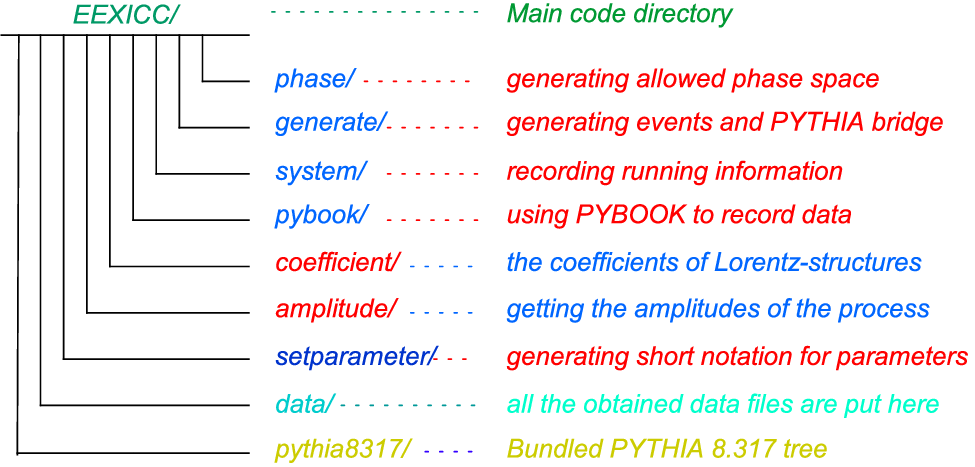}
\caption{The schematic structure for the generator EEXICC}
\label{struct}
\end{figure}

The schematic structure of the EEXICC generator is depicted in Fig.~\ref{struct}. The package contains seven physics and utility module directories, a data directory, and the bundled PYTHIA source tree. The generator is divided into seven modules according to their functions, and each module includes the essential files to complete the designated tasks for event generation. Additionally, the main directory {\tt EEXICC} contains three Fortran source files: {\tt parameter.F}, {\tt run.F}, and {\tt eexicc.F}.

\begin{itemize}

\item The module {\bf generate} is the core component of the program and consists of seven source files: {\tt evntinit.F}, {\tt genevnt.F}, {\tt pythia8\_io.F}, {\tt pythia8\_bridge.cc}, {\tt totfun.F}, {\tt initmixgrade.F}, and {\tt xiccpythia.F}. This module is responsible for initializing event-generation parameters, interfacing EEXICC with PYTHIA, evaluating the integrand for phase-space integration with the aid of the {\bf coefficient} and {\bf amplitude} modules, and performing the phase-space integration via the {\bf phase} module. In particular, the file {\tt initmixgrade.F} initializes the importance-sampling grid used in Monte Carlo simulations. Once the grid is generated by VEGAS, it can be stored and reused by {\tt initmixgrade.F} without rerunning the VEGAS integration. By appropriately setting the {\tt IMIX} and {\tt IMIXTYPE} parameters, different mixed-event configurations can be generated.

\item The module {\bf phase} contains three source files: {\tt phase\_gen.F}, {\tt phase\_point.F}, and {\tt vegas.F}. This module is responsible for generating physically allowed phase-space points and for recording the importance-sampling grid produced by VEGAS~\cite{Lepage:1977sw} into a grade file (with suffix \texttt{.grid}) stored in the {\it data} subdirectory. The file {\tt phase\_gen.F} implements a modified version of the RAMBOS algorithm~\cite{Kleiss:1985gy}, which generates multi-particle phase space and passes the resulting final-state four-momenta to the {\bf generate} module.

\item The module {\bf pybook} consists of five source files: {\tt pybookinit.F}, {\tt uphistrange.F}, {\tt uppydump.F}, {\tt uppyfact.F}, and {\tt uppyfill.F}. Its main purpose is to initialize and manage the PYBOOK subroutine in PYTHIA for recording generated events. This module can be optionally disabled in the main program file {\tt eexicc.F} if users prefer to implement alternative methods for event recording.

\item The module {\bf setparameter} includes two source files: {\tt simparameter.F} and {\tt uperror.F}. This module is designed to simplify and validate the input parameters defined in {\tt parameter.F}. If any input parameter lies outside its allowed range, predefined error messages provided by {\tt uperror.F} will be displayed, and the program execution will be terminated accordingly.

\item The module {\bf system} consists of six source files: {\tt upopenfile.F}, {\tt uplogo.F}, {\tt vegaslogo.F}, {\tt updatafile.F}, {\tt upclosegradefile.F}, and {\tt upclosepyfile.F}. This module handles the opening and closing of output files and displays runtime messages at various stages of program execution. These messages serve as progress indicators, informing users of the current execution status.

\item The module {\bf coefficient} contains seven source files: {\tt coefbb1s0.F}, {\tt coefbb3s1.F}, {\tt coefbc1s0.F}, {\tt coefbc3s1.F}, {\tt coefcc1s0.F}, {\tt coefcc3s1.F}, and {\tt coefee.F}. This module provides the numerical coefficients corresponding to each independent Lorentz structure appearing in the hard-scattering amplitudes.

\item The module {\bf amplitude} consists of four source files: {\tt ampy.F}, {\tt ampz0.F}, {\tt common.F}, and {\tt sqamp.F}. This module performs the numerical evaluation of scattering amplitudes using the Lorentz-structure coefficients supplied by the {\bf coefficient} module. The amplitudes for color $\mathbf{6}$ channels are obtained from the corresponding color $\overline{\mathbf{3}}$ channels by appropriately rescaling the color factors and the associated nonperturbative LDMEs.

\end{itemize}

Each module contains a {\bf makefile} used to build a library of the same name. For example, the {\bf makefile} in the {\bf generate} subdirectory is executed by GNU Make to produce {\tt generate.a} in the main directory. Once the executable \texttt{run} has been built, changing any of the parameters collected in \texttt{input.dat} does not require the program to be rebuilt: these parameters are read from \texttt{input.dat} at run time, and it is sufficient to execute \texttt{./run} again. A recompilation, i.e.\ a new invocation of \texttt{make}, is only needed when one of the Fortran source files or headers, or a compile-time parameter, has been modified; in that case only the affected files are recompiled. A master {\bf makefile} in the main directory orchestrates all sub-makefiles. Libraries required by the main program are declared in the master {\bf makefile}'s {\tt LIBS} variable and are built automatically via invocation of the sub-makefiles. This makefile-based implementation enhances the modularity and reusability of the EEXICC generator, enabling straightforward customization for diverse experimental scenarios.

All output files are systematically organized in the {\bf data} subdirectory, which contains twelve dedicated subdirectories for various doubly heavy baryon states. The file-naming convention follows a suffix-based classification: ``.grid'' files store importance-sampling functions, ``.cs'' files contain the adopted parameters and VEGAS runtime information, while ``.dat'' files record differential distributions, including the transverse-momentum and rapidity distributions of the doubly heavy baryons.

\subsection{Usage of EEXICC}

The EEXICC generator can be compiled by executing the \textbf{make} command in the root directory of the package. This produces an executable file named \textbf{run}. Model parameters can be specified either by directly modifying the source files (\texttt{run.F} and \texttt{parameter.F}) or, more conveniently, by editing the configuration file \texttt{input.dat}, which is designed to accommodate frequently adjusted parameters.

The main input parameters of EEXICC are summarized as follows:

\begin{itemize}

\item \texttt{PMB}: bottom-quark mass (in GeV); default value $5.10$~GeV. 

\item \texttt{PMC}: charm-quark mass (in GeV); default value $1.80$~GeV.

\item \texttt{PMZ}: $Z$-boson mass (in GeV); default value $91.1876$~GeV.

\item \texttt{PME}: electron mass (in GeV); default value $0.51\times10^{-3}$~GeV.

\item \texttt{ECM}: center-of-mass energy of the initial $e^+e^-$ system (in GeV); default value $91.1876$~GeV.

\item \texttt{SINTHETA2}: $\sin^2\theta_W$; default value $0.23119$.

\item \texttt{MSTU(111)}: specifies the perturbative order of $\alpha_s$ used in PYALPS. Setting \texttt{MSTU(111)}=0 keeps $\alpha_s$ fixed at the value defined by \texttt{PARU(111)}, while \texttt{MSTU(111)}=1 and 2 correspond to leading-order (LO) and next-to-leading-order (NLO) running, respectively.

\item \texttt{FULLDECAY}: total decay width of the $Z$ boson (in GeV); default value $2.4952$~GeV.

\item \texttt{IPROCESS}: selects the doubly heavy baryon to be generated. \texttt{IPROCESS}=1, 2, and 3 correspond to $\Xi_{bc}$, $\Xi_{cc}$, and $\Xi_{bb}$ production, respectively.
    
\item \texttt{IMIX}: controls whether mixed-event samples are generated. Setting \texttt{IMIX}=1 enables the mixing of different intermediate states, while \texttt{IMIX}=0 generates events from a single specified intermediate state.

\item \texttt{IMIXTYPE}: specifies the pattern of intermediate-state mixing. Four mixing modes are implemented:
\begin{itemize}
\item \texttt{IMIXTYPE}=1: all intermediate states for $\Xi_{bc}$ production;
\item \texttt{IMIXTYPE}=2: color $\overline{\mathbf{3}}$ states $(bc)[^1S_0^{\overline{\mathbf{3}}}]$ and $(bc)[^3S_1^{\overline{\mathbf{3}}}]$ for $\Xi_{bc}$ production;
\item \texttt{IMIXTYPE}=3: color $\mathbf{6}$ states $(bc)[^1S_0^{\mathbf{6}}]$ and $(bc)[^3S_1^{\mathbf{6}}]$ for $\Xi_{bc}$ production;
\item \texttt{IMIXTYPE}=4: the $(QQ)[^1S_0^{\mathbf{6}}]$ and $(QQ)[^3S_1^{\overline{\mathbf{3}}}]$ states for $\Xi_{cc}$ or $\Xi_{bb}$ production. 
\end{itemize}
This parameter only takes effect when \texttt{IMIX}=1.
    
\item \texttt{IXICCSTATE}: specifies the intermediate $(QQ')$ state to be generated. The options \texttt{IXICCSTATE}=1--4 correspond to
$(QQ')[^1S_0^{\overline{\mathbf{3}}}]$,
$(QQ')[^3S_1^{\overline{\mathbf{3}}}]$,
$(QQ')[^1S_0^{\mathbf{6}}]$, and
$(QQ')[^3S_1^{\mathbf{6}}]$, respectively. This parameter only takes effect when \texttt{IMIX}=0.

\item \texttt{NBOUND}: specifies the flavor of the light quark component in the produced doubly heavy baryon.
\begin{itemize}
\item \texttt{NBOUND} = 1: production of $\Xi_{QQ'u}$.
\item \texttt{NBOUND} = 2: production of $\Xi_{QQ'd}$.
\item \texttt{NBOUND} = 3: production of $\Xi_{QQ's}$.
\item \texttt{NBOUND} = 4: production of $\Xi_{QQ'}$, i.e., total cross section summed over the light flavors $u$, $d$, and $s$.
\end{itemize}
The default value is NBOUND = 1.
    
\item \texttt{FCC1SC3}, \texttt{FBC1SC3}, and \texttt{FBB1SC3}: radial wave functions at the origin, $R_{QQ'}(0)$, for color $\overline{\mathbf{3}}$ diquarks. By default, values are adopted from Ref.~\cite{Baranov:1995rc}. The corresponding color $\overline{\mathbf{3}}$ LDMEs are then calculated using Eq.~(\ref{eq.LDME3}).
    
\item \texttt{CMFACTOR}: the ratio between the color $\mathbf{6}$ LDMEs and the corresponding color $\overline{\mathbf{3}}$ LDMEs, i.e., $\langle{\cal O}^{\Xi_{QQ'}}(^{3(1)}S_{1(0)}^{\mathbf{6}})\rangle = \texttt{CMFACTOR}\cdot \langle{\cal O}^{\Xi_{QQ'}}(^{3(1)}S_{1(0)}^{\overline{\mathbf{3}}})\rangle$. The default value is 1.

\item \texttt{IVEGASOPEN}: controls the usage of the VEGAS algorithm. Setting \texttt{IVEGASOPEN}=1 activates VEGAS, while \texttt{IVEGASOPEN}=0 disables it.

\item \texttt{NUMOFEVENTS}: total number of events to be generated.

\item \texttt{MRPY(1)}: random seed, specified as \texttt{mrpy(1)} in \texttt{input.dat}. The default is 19780503 and the accepted integer range is $1\leq\texttt{mrpy(1)}\leq900000000$. The C++ bridge provides the hard-process random-number stream through \texttt{PYR}; a separate PYTHIA random-number generator receives the same seed. Distinct seeds should be used for separate samples, but do not constitute a proof of non-overlapping subsequences. For example, use seeds 12345 and 98765 in separate run directories, retaining the input card and integration settings for each sample. Regenerate grids when changing energy or physical parameters. Reproducibility also requires the same source, software environment and sequence of calls.

\item \texttt{IDWTUP}: PYTHIA parameter controlling the interpretation of event weights. For \texttt{IDWTUP}=3, all parton-level events are generated with unit weight and accepted without rejection. For \texttt{IDWTUP}=1, PYTHIA applies a hit-and-miss procedure (or VEGAS-based sampling) to unweight events, ensuring that all accepted events carry a common weight.

\item \texttt{IDPP}: determines the internal weight-handling strategy. When \texttt{IDPP}=1 (or 3), the event-weight treatment follows that of \texttt{IDWTUP}=1 (or 3), respectively. The option \texttt{IDPP}=2 generates unweighted events using an improved hit-and-miss algorithm, which significantly enhances efficiency while preserving the statistical accuracy of the cross-section evaluation.

\item \texttt{IGRADE}: determines whether an existing VEGAS integration grid is reused. Setting \texttt{IGRADE}=1 enables the reuse of a previously generated grid (stored in a \texttt{.grid} file), significantly reducing computation time when relevant parameters remain unchanged. Setting \texttt{IGRADE}=0 forces a new grid generation.

\end{itemize}

EEXICC offers three Monte Carlo simulation strategies: a standard method without VEGAS, and two VEGAS-based importance-sampling approaches. For the latter, event generation can utilize either a pre-existing integration grid from a previous run or a newly generated grid. By setting \texttt{IVEGASOPEN}=0 and \texttt{IGRADE}=1, the generator employs an existing importance-sampling function -- an adaptive probability distribution optimized by VEGAS to concentrate sampling on high-contribution regions of the multi-dimensional phase space -- enabling efficient event generation without rerunning the VEGAS integration. Once an appropriate grid is obtained, VEGAS needs to be executed only once unless parameters affecting the integration behavior (e.g., phase-space boundaries, hard-scattering dynamics) are modified.

For theoretical analyses, such as precise calculations of production cross sections and differential kinematic distributions, weighted events are sufficient and computationally optimal. In EEXICC, the most efficient configuration for such studies is to set \texttt{IDPP}=3 (equivalent to the PYTHIA parameter \texttt{IDWTUP}=3) or \texttt{IDPP}=1 (equivalent to \texttt{IDWTUP}=1) to generate weighted events for the target doubly heavy baryons. As a subtle point, in these theoretical calculation scenarios, the parameter \texttt{XMAXUP} should be set to 0 to ensure correct normalization of cross sections. Conversely, for realistic detector simulations that aim to reproduce experimental observables, response functions, and detector efficiencies, unweighted events are strictly required, as each event must carry a uniform physical weight to avoid bias in the simulated data. In EEXICC, such standard unweighted events are generated by setting \texttt{IDPP}=1 or 2, which employ appropriate hit-and-miss or optimized sampling algorithms to produce physically consistent event samples for experimental analyses.

\subsection{A test run}

\begin{figure}[htb]
\centering
\includegraphics[width=0.48\textwidth]{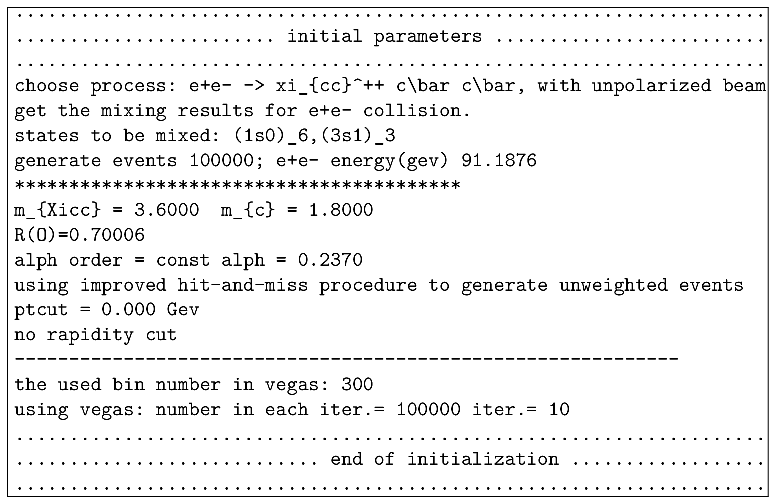}
\caption{Snapshot of the initial parameters used in the test run of the EEXICC generator, which generates $10^5$ unweighted $\Xi_{cc}^{++}$ baryon events for the mixed states $(cc)[^1S_0^{\mathbf{6}}]$ and $(cc)[^3S_1^{\overline{\mathbf{3}}}]$.}
\label{test}
\end{figure}

A test run was performed to simulate the production of $\Xi_{cc}^{++}$ events at an $e^+e^-$ collider running at the $Z$ pole. Unweighted events were generated using the hit-and-miss method by setting $\texttt{IDPP}=2$, with the input parameters displayed in Fig.~\ref{test}. The parameter snapshot shown in Fig.~\ref{test} is reproduced by the following steps:
(1) Run \texttt{make} in the main directory;
(2) edit \texttt{input.dat} such that \texttt{ECM=91.1876}, \texttt{IPROCESS=2}, \texttt{IMIX=1}, \texttt{IMIXTYPE=4}, \texttt{NBOUND=1}, \texttt{IDPP=2}, \texttt{NUMOFEVENTS=100000}; 
(3) run \texttt{./run}. 

The generated data were compressed into a file named \texttt{testdata.zip}, located in the main directory of the program.

With the parameters shown in Fig.~\ref{test}, a reasonably accurate production cross section can be obtained. To acquire precise differential distributions, the number of generated events and the VEGAS parameter ``ncall~\footnote{\texttt{ncall} is defined in the header file \texttt{invegas.h}.}'' should be increased to $10^7$ and $10^6$, respectively. Under these conditions, we obtain the cross section
\begin{eqnarray}
\sigma(\Xi_{cc}^{++}) = 0.336\,\text{pb}.
\end{eqnarray}
Differential distributions for $\Xi_{cc}^{++}$ production, specifically the polar angle ($\cos\theta$)~\footnote{The angle $\theta$ is defined as the angle between the momenta of the initial electron and the produced $\Xi_{cc}^{++}$ in the center-of-mass frame of the initial $e^+e^-$ system.}, energy fraction ($z$)~\footnote{Here, the energy fraction is defined as $z=2p_1 \cdot (q_1+q_2)/(q_1+q_2)^2$.}, transverse momentum ($p_{t}$), and rapidity ($y$), are presented in Fig.~\ref{distribution}. This figure explicitly shows the contributions from the two intermediate diquark states: $(cc)[^3S_1^{\overline{\mathbf{3}}}]$ and $(cc)[^1S_0^{\mathbf{6}}]$.

\begin{widetext}
\begin{center}
\begin{figure}[htb]
\includegraphics[width=0.85\textwidth]{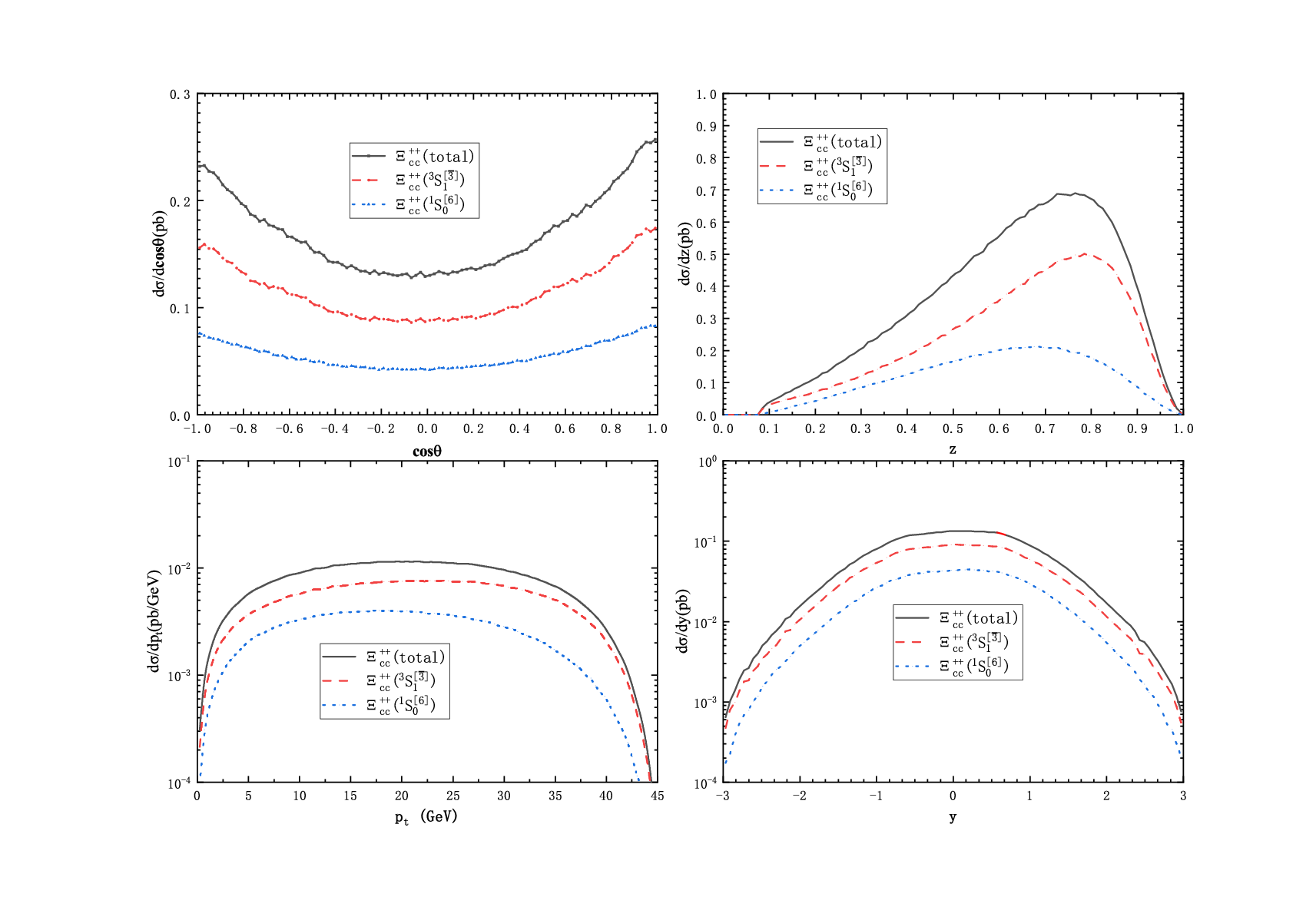}
\caption{Differential distributions of ${\rm cos}\theta$, $z$, $p_{t}$, and $y$ for the test run of $\Xi_{cc}^{++}$ production.} \label{distribution}
\end{figure}
\end{center}
\end{widetext}

\section{Summary}

Future high-luminosity $e^+e^-$ colliders running around the $Z$ pole, such as the CEPC and FCC-ee, provide a clean experimental environment that complements hadron colliders for studying doubly heavy baryons ($\Xi_{cc}$, $\Xi_{bc}$, and $\Xi_{bb}$). To facilitate precision studies at these facilities, we have developed EEXICC, a dedicated Monte Carlo event generator designed to simulate the production of doubly heavy baryons via $e^+e^-$ annihilation.

EEXICC uses modular Fortran routines for the hard process and a C++ interface to PYTHIA 8.317. To overcome the computational complexity inherent in multi-body final states, we implement the ``improved trace technique,'' which performs calculations at the amplitude level to significantly enhance numerical efficiency. Additionally, the program incorporates the VEGAS algorithm for optimized phase-space integration and supports the generation of unweighted events. Validated against existing theoretical results, EEXICC serves as a robust and flexible tool for investigating the properties of doubly heavy baryons at next-generation lepton colliders.

\hspace{2cm}

\noindent {\bf Acknowledgments:}  This work was supported in part by the Natural Science Foundation of China under Grants No.12575080, No.12547101 and No.12547115, and by the Chongqing Natural Science Foundation under Grant No. CSTB2025NSCQ-GPX0745.

\end{document}